\documentclass[prc,aps,showpacs,floatfix,nofootinbib,preprint]{revtex4} 
\pdfoutput=1
\usepackage{graphicx} 
\usepackage{amssymb} 
\usepackage{amsmath} 
\usepackage{amsfonts}

\def\be{\begin{equation}} 
\def\ee{\end{equation}} 
\begin{document} 
\title{Estimating parameter uncertainty in binding-energy models by the
frequency-domain bootstrap.}
\author{G.F. Bertsch$^{1}$ and Derek Bingham$^{2}$}
\affiliation{$^{1}$Department of Physics and Institute of Nuclear Theory, 
Box 351560\\ University of Washington, Seattle, Washington 98915, USA\\
$^{2}$Department of Statistics, Simon Fraser University, Vancouver, CA} 
 
\begin{abstract} 
 
We propose using the frequency-domain bootstrap (FDB) to estimate errors 
of modeling parameters when the modeling error is itself 
a major source of uncertainty.  Unlike the usual bootstrap or the
simple $\chi^2$ analysis, the FDB can take into account correlations between
errors.  It is also very fast compared to the
the Gaussian process Bayesian estimate
as often implemented for computer model calibration.  The method is illustrated with a simple example, the liquid
drop model of nuclear binding energies.  We find that the FDB gives
a more conservative estimate of the uncertainty in liquid drop parameters
in better accord with more empirical estimates.  
For the nuclear physics application, there 
no apparent obstacle to apply the method
to the more accurate and detailed models based on density-functional theory.

\end{abstract}

\maketitle

{\it Introduction.} The {\em bootstrap} method is 
widely used to estimate sampling distributions of statistics
\cite{ef77}.   We will show
here that the frequency-domain bootstrap  (FDB) for time series analysis
\cite[Sect. 6]{kr12} is
well suited for estimating uncertainty in the modeling parameters 
arising in the theory of nuclear binding energies.  Parameters such
as the binding energy of nuclear matter and symmetry energy of nuclear
matter are needed to construct models of the nuclear equation of state,
which is an essential ingredient in the physics of neutron stars.
We first describe the method in general terms, and then apply it to the very simple liquid-drop model of
nuclear binding energies.  The same approach can be applied to more
sophisticated models, such those based on density-functional theory
\cite{re16}, 
which should provide even narrower limits than can be obtained from the liquid drop
description.

{\it $\chi^2$ and the basic bootstrap.}  Estimating parameters 
in models  and their respective uncertainties is fraught with difficulty.  Obviously,
if there is theoretical guidance on the functional form of the systematic difference between the
system response and the model, it should be incorporated into the parameter estimation.
Absent any guidance, the estimation can on be based on the model's
performance; parameters that make the model ``look" like data are preferred.

Denote the model function $M(x,p)$; it depends on
parameters $p$ and maps input data variables
$x$ (which may be vectors) onto output  $y$.  For the
nuclear physics model treated below, the variables are $x = (Z,N)$,
the proton and neutron numbers, and $y=M(x,p)$ is the binding
energy of the nucleus.  

The first step in applying a model is to determine a best-fit parameter vector, $p_0$, by 
minimizing the square of the residual differences $r$ between the model prediction and
the experimental data,  $r(x) =  y_{exp}(x) - M(x,p)$.  Denote the corresponding
vector of the best-fit residuals by $r_0$.   The experimental data for this sort of procedure 
is implicitly specified as 
\be
y_{exp} = M(x, p) + C(x)
\ee
where $C(x)$ is a correction or error term.  
The perfect model would have
the form $y_{exp} = M(x,p_t) + C(x)$, where $p_t$ are the true parameters and $r(x)=C(x)$.  

Now comes the main assumption of the
$\chi^2$ method:
The correction terms, $C(x)$, are independent for each $x_i$ ($i=1,\ldots,S$) and follow a mean zero Gaussian distribution with equal variances (the equal variance assumption specifies that the experiments have the same uncertainty and is not strictly necessary).  The likelihood for the parameters is then
\be
L(p) =\frac{1}{(2\pi \sigma_0^2)^{S/2}} \exp(-\sum_{i=1}^S (y_{exp}(x_i) - M(x_i,p) )^2/ 2 \sigma_0^2),
\ee
where $\sigma_0^2$ is the residual variance.  

Broadly speaking, the bootstrap is an approach to estimating the sampling distribution of a statistic that requires few assumptions on the 
process that generated the data.  The basic idea is that the data, or some function of the data, is repeatedly sampled with replacement.
For each bootstrap sample, parameters of interest are estimated, and the ensemble of parameter estimates is the corresponding bootstrap distribution.
The basic bootstrap in our setting can now be defined as an approximation to sampling distribution of the estimator of $p_t$ without specifying a particular error distribution.   To do so, we (i) draw samples, with replacement, of residuals from the entries in the $r_0$ vector and (ii) reestimate $p_t$.  Doing this many times gives the bootstrap distribution of $p_t$ values.  From this bootstrap distribution, functionals such as uncertainty estimates or confidence intervals can be computed.

 {\it Dealing with correlations.}  The $\chi^2$, or basic bootstrap method for that matter, 
greatly underestimates the uncertainty of the parameter distribution in
many circumstances.  The reason is that the assumed ensemble has no
correlation between the residuals at different $x$ points; an unlikely occurrence because 
if the model overestimates the system mean response at a given $x$, it is also likely to overestimate the mean at nearby $x$'s as well.  
If the residuals are correlated,
that needs to be taken into account in constructing the sampling distribution
function for the estimator of $p_t$, otherwise the variance
in the derived sampling distribution will be too small.  

Correlations can be taken into account by a 
Gaussian process ensemble of residuals \cite{ke01} and this method has
become an accepted tool in nuclear physics \cite{mc15,hi15,pr15,be16} and 
elsewhere in physics \cite{recent} under the heading of computer model calibration.  This amounts to an attempt to consider any systematic
signal in the residuals, as a function of $x$, that is not accounted for by the model. The specification a Gaussian process requires that a mean function (usually taken to be a constant) and a correlation function must be chosen.  

The specification in (1) is essentially the same
as adopted in  \cite{ke01} where their correction term is the sum of a discrepancy function and experimental error.
In the applications we consider, the experimental error is a negligible component of $C(x)$ and can be
safely ignored.  Therefore, $C(x)$ is analogous to their discrepancy function.
The frequency-domain bootstrap that we advocate here is an alternative
way to include correlations.  Furthermore, 
it is computationally much more efficient to the Gaussian process approaches in  \cite{ke01,hi15}.

{\it The Frequency-Domain Bootstrap.} 
As before, we start with
a parameter fit $p_0$ producing a residual vector $r_0$. We need to have
a measure of distance between data points, $|x_i-x_j|$, and we assume
for the moment that $x$ is one-dimensional array of contiguous
integers.  This allows us to define the discrete Fourier transform
$\tilde r$ of the residual vector $r$.  For a residual vector of 
dimension $S$, the discrete Fourier transform (and its inverse) may be
expressed
\be
\tilde r(m) = S^{-1/2} \sum_n \exp(2 \pi i n m/ S) r(n),
\ee
with $m$ in the range $-S/2 < m < S/2$.  The Fourier transform
can also be expressed with pure real variables as
\be
r(n) = \tilde r(0) +\left({2\over S}\right)^{1/2}\sum_{m=1}^{S/2} |\tilde r(m)| 
\sin (2 \pi i n m /S + \phi_m).
\ee
If there are strong correlations of short range in $x$, the magnitudes of
the residuals will be enhanced for $m << S/2$.  On the other hand, if there
are no correlations between difference $x$ points we expect the components
of $\tilde r(m)$ to be Gaussian distributed with a variance independent of
$m$.  We have no information about the phases, $\phi_m$, and we assume that
they are uniformly distributed in the interval $0 < \phi < 2\pi$ to
construct the FDB ensemble.  For the ``bootstrap" ensemble, we sample, with replacement,
the values of $|\tilde r(m)|$ from the $r_0$ set of residuals.  The
uncertainties are calculated as before: (i) sample the ensemble; (ii) refit
the model to get a sample $p_t$; and (iii) extract the statistical
uncertainty by the variance of the $p_t$ samples.

Similar to the Gaussian process ensemble approach of Kenney and O'Hagan 
\cite{ke01},
our approach has an implicit Gaussian assumption for $C(x)$.  
The main difference is that they view the correction as a realization of a stationary Gaussian process with a specifically chosen
covariance function to relate the values of $C(x)$
at inputs $x_i$ and $x_j$.  While our approach also assumes an underlying Gaussian process, we do not have to specify
the correlation function.  In  \cite{hi15}, the chosen correlation function specifies that $C(x)$ is infinitely differentiable.
The FDB does not make this smoothness assumption. In this sense, the proposed it is more flexible and contains the
specification in \cite{hi15} as a special case.
Furthermore, the Gaussian process approach adopted by \cite{ke01, hi15}  requires
the inversion of an $S \times S$ correlation matrix for each evaluation of the Gaussian likelihood.  To implement
their method can require tens-of-thousands of such inversions, each with order $o(S^3)$ \cite{ka11}.
For the liquid drop model in the next section, the sample size, $S$ is in the thousands, thereby making their
approach less computationally appealing.  The proposed approach, on the other hand, makes computations in the order of $o(S\log(S))$.

{\it Application to the liquid drop model.}  
A standard formulation of the liquid drop model of nuclear binding energies
$B$ is
\be
B(Z,N) = a_v A - a_s A^{2/3} - a_c {Z^2\over A^{1/3}} 
-a_a {(N-Z)^2\over A} -\delta { {\rm mod}(Z,2) + {\rm mod}(N,2) -1\over A^{1/2}},
\label{LDE}
\ee
where $Z$ and $N$ are the proton and neutron numbers respectively, and $A=Z+N$.
The coeffients of the terms in Eq. (\ref{LDE}) have clear physical
interpretations. The nuclear matter binding energy $a_v$ and the asymmetry
term $a_a$ are the only ones to
survive in the nuclear matter limit, providing the Coulomb energy term
$a_c$ is externally compensated (as in a neutron star).  It should be
emphasized that this should be considered a toy model for the physical
problem due to the omission of shell effects.  They are included in
models based on nuclear energy density functionals; those models achieve
a factor of two better accuracy at a cost of a factor of two in the
parameter count \cite{be05}.

We first determine the $p_0$ parameter set by least-squares 
minimization of the residuals.  The data
set is the experimental binding energies of 2037 nuclei from the 2003 nuclear
mass table \cite{audi}.  The resulting fit gives $a_v=15.58 $ MeV and
$a_a = 22.18$ MeV, with a variance of the binding energy residuals
$\sigma_r = 3.24$ MeV.  The nuclear matter binding energy has hardly
changed by the additional data of the last 50 years.  The first fit
that included an error estimate \cite{my66} found $a_v=15.68 $ MeV.  Their
error estimate was ``say to 1 or 2 \% ."

We first carry out the $\chi^2$ estimate of the parameter uncertainties,
with the result for $a_v$ shown as the top line in Table I.  
We next carry out the basic bootstrap, 
randomly assigning the residuals to 
different $(Z,N)$ and re-optimizing the parameters.  The results, 
shown on the second line of the Table, confirm  that the method
is a good approximation to $\chi^2$.
\begin{table}[htb] 
\begin{center} 
\begin{tabular}{|c|cccc|cc|} 
\hline 
 Treatment  & $\chi^2$ &   BB  &  FDB  &  GP   &  Ref. \cite{ma17}  & 
 Ref. \cite{re16}  \\
\hline 
$\bar a_v$   &  15.58 & 15.58 & 15.56 & 15.58 & 15.88 & 15.81 \\ 
$\sigma_{a_v}$  & 0.03 &  0.03 & 0.17 & 0.12 & 0.15 & 0.17 \\ 
\hline 
\end{tabular} 
\caption{Nuclear binding energies and their uncertainties estimated by different statistical
methods.  The first four entries show the results for the liquid drop
model:
$\chi^2$, the standard chi-squared;  BB, the basic bootstrap;
FDB, the frequency-domain bootstrap; and GP, the Gaussian process method.
For comparison, the last two entries show estimates 
based on the Skyrme family of energy density functionals.
Energies are in units of MeV.
}
\label{compare} 
\end{center} 
\end{table} 
However, the estimated uncertainty, 
$\sigma_{a_v} = 0.03 $, is wildly unrealistic.  This may be seen from
alternative formulations of the liquid drop model \cite{Z2} or from
improved models that include more of the actual physics.  Examples
using parametered energy-density
functionals are on the bottom two rows of the Table.  The quoted
uncertainties are 5 times larger than the naive estimates in the
Table.

The problem of course is that the ordinary bootstrap assumes that the
residuals are uncorrelated--true for experimental data but not for
model errors.  This may be seen in Fig.~1 showing the nuclei in the
data set with the sign of $r(Z,N)$ indicated by color.  
\begin{figure}[tb] 
\begin{center} 
\includegraphics[width=8 cm]{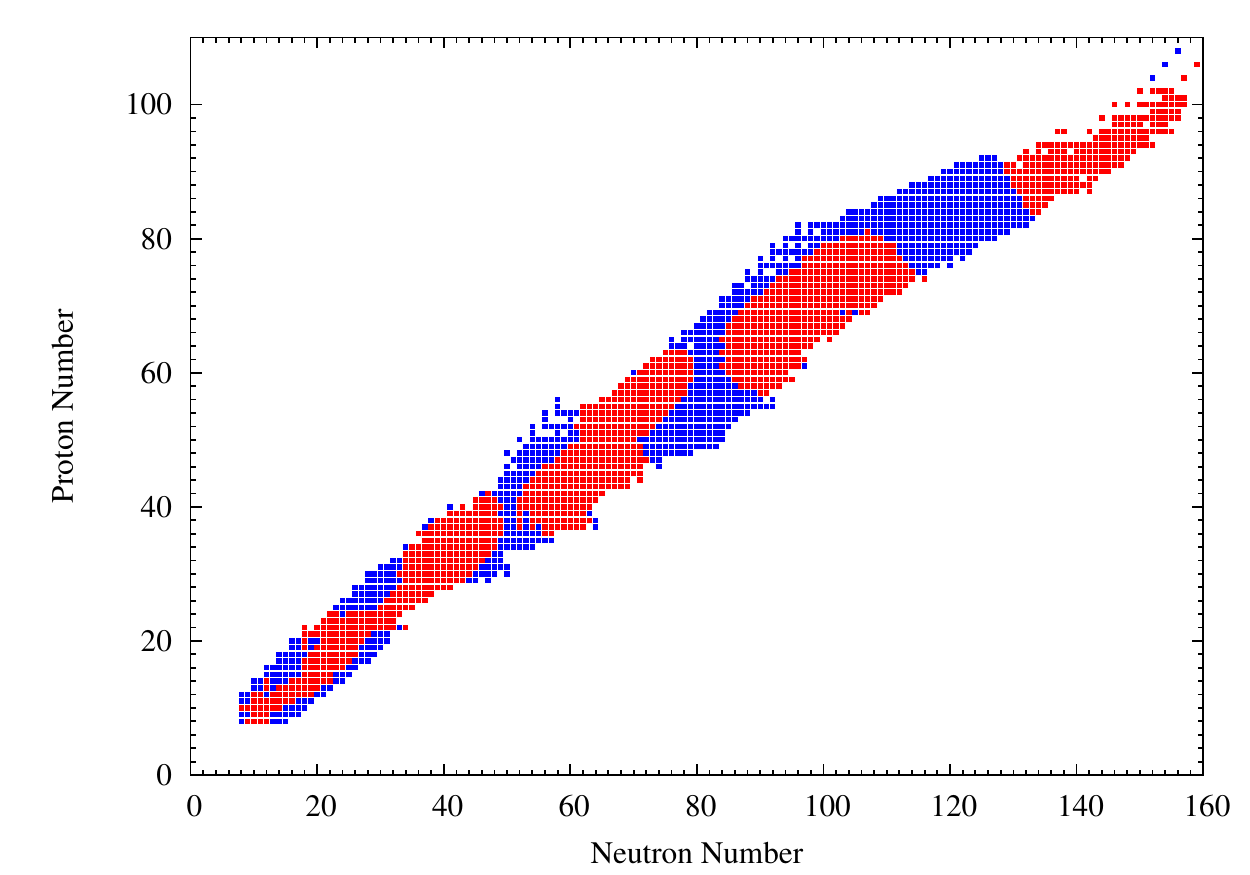}
\caption{Sign of the error in the liquid drop fit to nuclear binding
energies.  Blue: model binding too high; red: model binding too low.
}\
\label{color} 
\end{center} 
\end{figure} 
More quantitatively,
Fig. \ref{residA} shows the residuals as a function of $A$, averaging over
the nuclei in the data set with given $A$.  
\begin{figure}[tb] 
\begin{center} 
\includegraphics[width=8 cm]{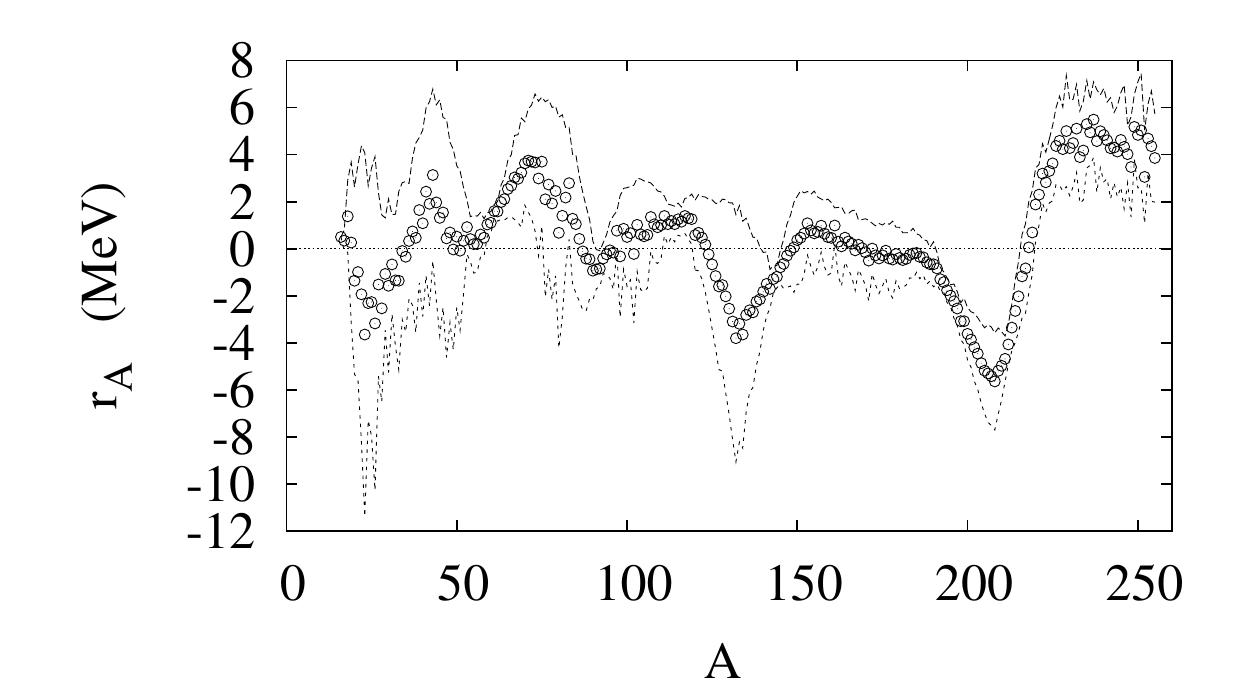} 
\caption{Distribution of residuals as a function of $A$.  See text for
explanation}
\label{residA} 
\end{center} 
\end{figure} 
The central circles are the average residual 
of fixed $A$, $r_A = \sum_{Z+N=A} r(Z,N)/N_A$ where $N_A$ in the number of
data.  The two curves delimit the variance of the residuals for a 
fixed $A$.  Obviously, the residuals are obviously highly correlated.

\begin{figure}[tb] 
\begin{center} 
\includegraphics[width=8 cm]{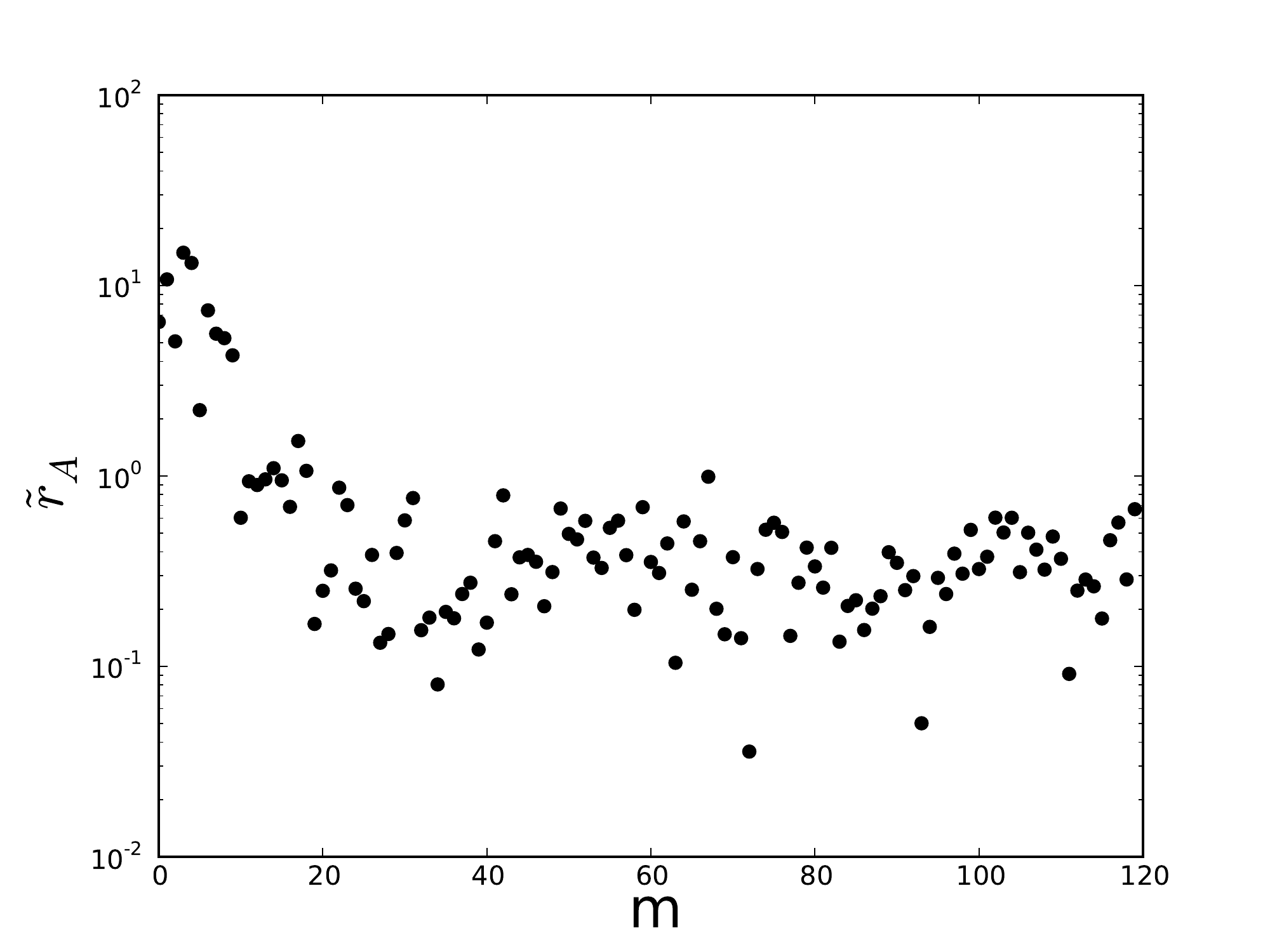}
\caption{Distribution of $|\tilde r(n)|$ for the FDB.}
\label{rq} 
\end{center} 
\end{figure}

\begin{figure}[tb] 
\begin{center} 
\includegraphics[width=8 cm]{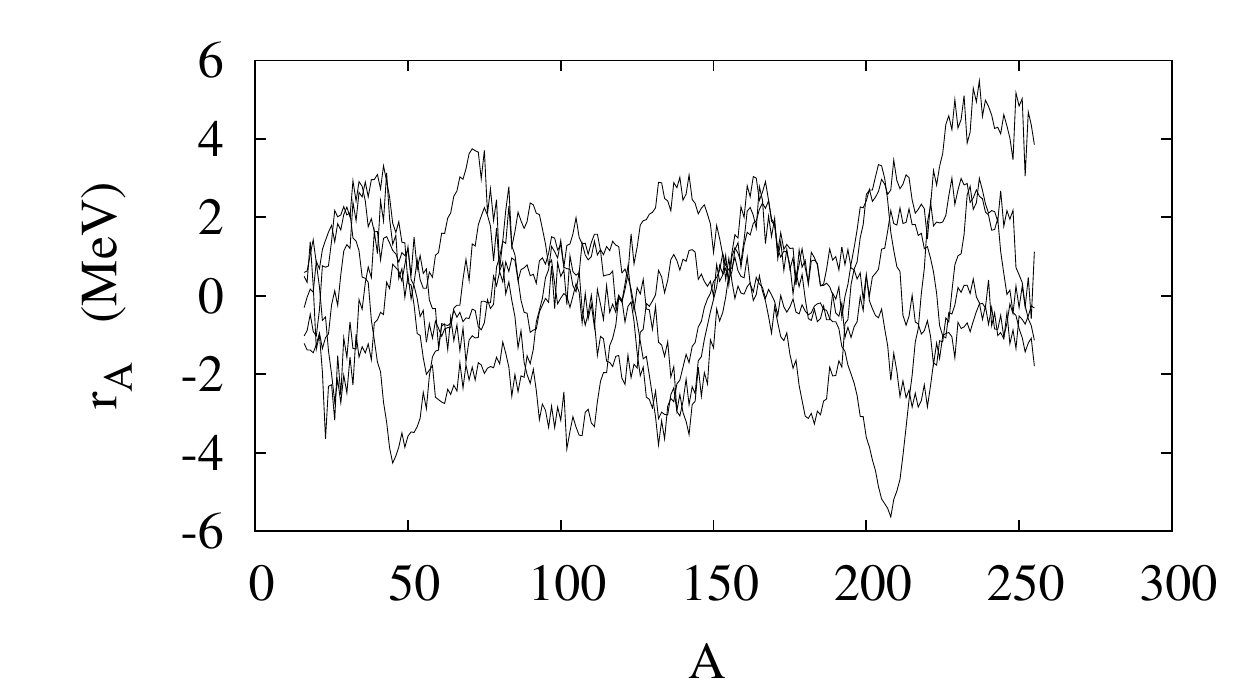}
\caption{Sample $r_A$ sets for the liquid drop model obtained by the
FDB method}
\label{samples} 
\end{center} 
\end{figure}

\begin{figure}[tb] 
\begin{center} 
\includegraphics[width=8 cm]{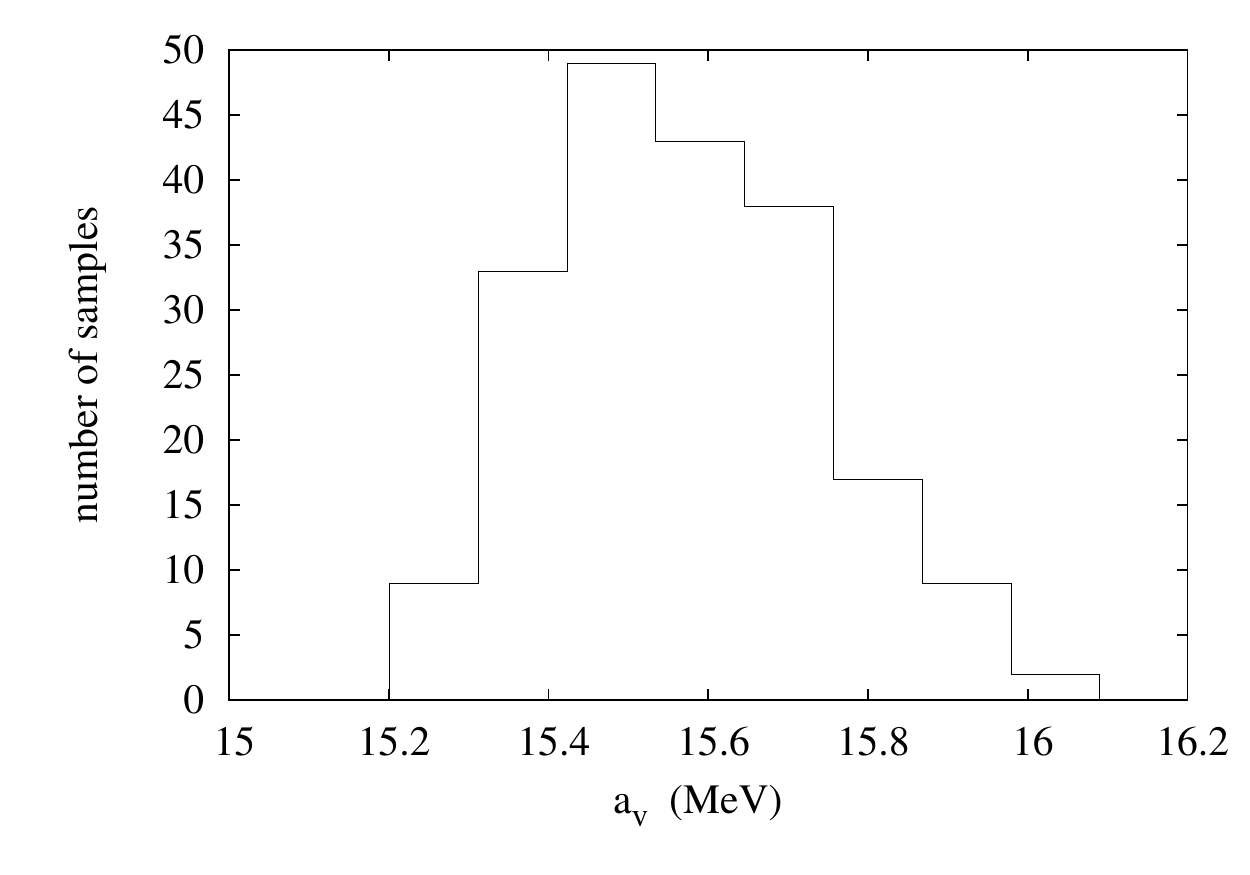}
\caption{Distribution of $a_v$ obtained by the
FDB method}
\label{fdb_av} 
\end{center} 
\end{figure}

We now construct the FDB ensemble.  We take $A$ to be the variable in
the Fourier transform and $r_A$ as the $r_0$ data set.  The
Fourier-transformed residuals $|\tilde r_A|$ are plotted in Fig. \ref{rq}.
We take samples of the $r_A$ distribution by inverse Fourier transforming
$|\tilde r_A| \exp{i \phi_A}$ where $\phi_A$ is chosen to be uniform in
the interval $[0,2\pi]$.  
Several
samples are shown in Fig. \ref{samples}.   We see that the locations of 
the peaks and troughs can be different from sample to sample.
Before we can refit within our new ensemble we have to decide
how to deal with the dependence of the full residual
function $r(Z,N)$ on $N-Z$.  The problem of correlations is less severe
here because, as may be seen in Fig. 1, the chains of nuclides in the $N-Z$ direction are 
quite short.  We choose to deal with
the $N-Z$ degree of freedom by taking a
$\chi^2$ distribution
of residuals $r(Z,N)$ about $r_A$.  The variance of the distribution
is taken from the $r_0$ data set, 
\be
\sigma_A^2 = 
{1\over N_A } \left(\sum_{Z+N=A} r(Z,N)^2 - r_A^2\right). 
\ee
Following this procedure, we generated parameter sets from 200 samples.
A histogram of $a_v$ values is shown in Fig. \ref{fdb_av}.
The distribution is much broader than that of the ordinary bootstrap.  
The variance
in $a_v$ values comes out to be $\sigma_{a_v} = 0.17 $ MeV, which
appears to be quite reasonable compared with the results from the
better models quoted in the last columns  of Table I.  

{\it Comparison with the GP method.}  As mentioned earlier, ensembles
of residuals based on the Gaussian process have become very popular.
We have implemeted the fully Bayesian estimate outlined in \cite{hi04} 
to estimate parameter
uncertainties using a 5-parameter Gaussian process as part of the
distribution, with results shown on the fourth line of the Table.
The uncertainty $a_v$ is somewhat smaller than the FDB estimate. That is
perhaps to be expected. The Fourier transform can capture
many degrees of freedom for the correlations, while the GP is limited
to the number of parameters in the Bayesian ensemble.   Put another way,
the methods outlined in  \cite{ke01,hi04} posits a specific type of correlation structure
for the function $C(x)$, while the FDB considers a broader class of Gaussian process models.
The result of the increased flexibility is additional uncertainty in the estimated form of $C(x)$.
Whether or not this is a good thing or not depends on how strongly one
believes in the more restrictive choice of correlation function in \cite{ke01, hi04}.

{\it Discussion. }  We have demonstrated that the FDB gives a better
estimate of parameter uncertainty than the $\chi^2$ method,
which is very well known to unrealistic in the description of nuclear
properties \cite{do14}.  Certainly, the uncertainties obtained via the FDB 
attempt to capture the correlations that the standard $\chi^2$ method
ignores.  Whether the FDB estimate is large enough to be realistic
can still be questioned.  We have compared with the results from one
family of density functional models, but other models can give larger
deviations.  Broadly speaking, the advantages of the FDB 
approach are computational and a
broader exploration of correlation models than methods in \cite{ke01, hi04}.
We expect to generally encounter more parameter
uncertainty as a result of this broader exploration.

{\it Acknowledgment.}  This work was stimulated by the program
``Bayesian methods in nuclear physics" at the Institute for Nuclear Theory
at University of Washington.  GB also acknowledges helpful discussion with
W. Nazarewicz and J. Margueron.  The research was partially funded
by the Natural Sciences and Engineering Research Council of
Canada.


\begin{thebibliography}{99} 
\bibitem{ef77} B. Efron, Annals of Statistics {\bf 7} 1 (1979). 
\bibitem{kr12}  J-P. Kreiss and S.N. Lahiri, Handbook of Statistics
{\bf 30} 3 (2012).
\bibitem{re16} P.G. Reinhard and W. Nazarewicz, Phys. Rev.C {\bf 93}, 051303  (2016)
\bibitem{ke01} M.C. Kennedy and A. O'Hagan, J. R. Statist. Soc. B
{\bf 63} 425 (2001).
\bibitem{mc15} J.D. McDonnell, N. Schunck, D. Higdon, J. Sarich, S.M.
Wild, and W. Nazarewics, Phys. Rev. Lett. {\bf 114} 122501 (2015).
\bibitem{hi15} D. Higdon, D. MacDonnell, N. Schunck, J. Sarich and
S. Wild, J. Phys. G {\bf 42} 034009 (2015).
\bibitem{pr15} S. Pratt, et al., Phys. Rev. Lett. {\bf 114} 202301 (2015)
\bibitem{be16} J. Bernard, et al., arXiv:1605.0395 (2016).
\bibitem{recent} Two recent examples are:  C. Moore, et al., Phys. Rev.
D {\bf 93} 064001 (2016); J. Cui and R. Krems, Phys. Rev. Lett. {bf 115}
073202 (2015). 
\bibitem{ka11} C.G. Kaufman , D. Bingham, S. Habib, K. Heitmann, J.A.
Frieman, Ann. Appl. Stat. {\bf 5} 4 (2011).
\bibitem{be05} G.F. Bertsch, B. Sabbey and M. Uusnakki, Phys. Rev. C{\bf 71}
 054311 (2005). 
\bibitem{audi}G. Audi, A. H. Wapstra, and C. Thibault, Nucl. Phys.
{\bf A729} , 337 (2003).  Nuclei were selected for the fit by the
criteria $Z>7$, $N>7$, $A<256$, and experimental uncertainty on
binding energy less than 0.2 MeV.
\bibitem{my66} W. Myers and W. Swiatecki, Nucl. Phys. {\bf 81} 1 (1966).
\bibitem{ma17}  J. Margueron, R. Casali, and F. Gulminelli, to be published.
\bibitem{Z2} For example, the replacement $Z^2 \rightarrow Z(Z-1)$ in 
Eq. (\ref{LDE}).
\bibitem{hi04} D. Higdon, SIAM J Sc. Comput. {\bf 26} 448 (2004).
\bibitem{do14} J. Dobaczewski, W. Nazarewicz, and P-G. Reinhard,
J. Phys. G {\bf 41} 074001 (2014).
\end{thebibliography}
\end{document}